\documentclass[pra,superscriptaddress,twocolumn]{revtex4-2}

\usepackage{enumerate}
\usepackage{mathtools}
\usepackage{amsfonts,amssymb,amsmath}
\usepackage[]{graphics,graphicx,epsfig}
\usepackage{amsthm}
\usepackage{graphicx}
\usepackage{dcolumn}
\usepackage{natbib}
\usepackage{color}
\usepackage{multirow}
\usepackage{ulem}
\usepackage{bm}
\usepackage{url}

\newcommand{\ket}[1]{|{#1}\rangle}

\usepackage{xcolor}

\begin{document}


\title{Sending absolutely maximally entangled states through noisy quantum channels}

\date{\today}

\author{Maria Stawska}
\affiliation{Institute of Spintronics and Quantum Information, Faculty of Physics and Astronomy, Adam Mickiewicz University, 61-614 Poznań, Poland}

\author{Jan W\'{o}jcik}
\affiliation{Institute of Spintronics and Quantum Information, Faculty of Physics and Astronomy, Adam Mickiewicz University, 61-614 Poznań, Poland}

\author{Andrzej Grudka}
\affiliation{Institute of Spintronics and Quantum Information, Faculty of Physics and Astronomy, Adam Mickiewicz University, 61-614 Poznań, Poland}

\author{Antoni W\'{o}jcik}
\affiliation{Institute of Spintronics and Quantum Information, Faculty of Physics and Astronomy, Adam Mickiewicz University, 61-614 Poznań, Poland}


\begin{abstract}
Absolutely maximally entangled states are quantum states that exhibit maximal entanglement across any bipartition, making them valuable for applications. This study investigates the behavior of qubit AME states under the influence of noisy quantum channels. Our results demonstrate that for certain channels, such as the depolarizing channel, the entanglement properties remain invariant under local unitary transformations and are independent of the choice of qubits in each subset. However, for channels like the dephasing channel, the entanglement behavior can vary depending on the specific AME state and the choice of qubits, revealing a symmetry-breaking effect. These findings highlight the nuanced relationship between AME states and noise, providing insights into their robustness and potential applications in noisy quantum systems.
\end{abstract}

\maketitle

\section{Introduction.} Absolutely Maximally Entangled (AME) states are among the most interesting many-particle quantum states~\cite{PhysRevA.86.052335,helwig2013absolutely}. These states are maximally entangled between any two disjoint subsets of qudits, existing only for specific combinations of dimensions and numbers of qudits. Their most important applications include the construction of holographic states used as toy models for AdS/CFT correspondence~\cite{pastawski2015holographic}, as well as quantum communication protocols like parallel teleportation \cite{PhysRevA.86.052335} and quantum secret sharing \cite{PhysRevA.86.052335}.

In practical implementations, we must deal with mixed states rather than pure states due to noise in both preparation and distribution processes. The effect of noise on various multipartite states has been extensively studied~\cite{PhysRevA.65.052327,PhysRevLett.92.180403,PhysRevA.71.032350,PhysRevA.72.042339,PhysRevA.78.064301,PhysRevA.82.032317}. While noise typically reduces entanglement, entanglement distillation protocols~\cite{PhysRevLett.76.722,PhysRevLett.77.2818,PhysRevA.54.3824} can sometimes extract pure entangled states from mixed ones, which is particularly important for quantum communication tasks.

In this work, we examine the action of various quantum channels on qubit AME states, focusing on symmetric channels where each qubit undergoes identical noise. While the entanglement necessarily decreases, we investigate whether the initial permutation symmetry of the AME state's entanglement measures persists. It is important to note that while AME states need not be permutationally invariant themselves, their entanglement measures must be permutationally invariant. We demonstrate that some AME states maintain this symmetry under noise while others exhibit symmetry breaking.

To quantify this behavior, we compute logarithmic negativity (which is an upper bound on destillable entnglement) and coherent information (which is a lower bound on destillable entanglement) for all partitions of qubits. 

\section{Model} Let us consider the set of $n$ qudits ($d$ dimensional quantum systems) combined in pure state $|\Phi\rangle$. This state can be always split into two parts $A$ (with $m$ qudits) and $B$ (with $n-m$ qudits). Without losing generality let us also assume that $m\leq n/2$. Density matrices for each subset can be obtained by tracing over the other one
\begin{equation}
    \rho_A = \text{Tr}_B |\Phi\rangle\langle\Phi|.
\end{equation}
Due to purity of the state $|\Phi\rangle$ for a given partition $(A/B)$ entanglement between subsystems $A$ and $B$ can be obtained from von Neumann entropy 
\begin{equation}
    S(\rho_A) = -\text{Tr}(\rho_A \text{log}\rho_A).
\end{equation}
Note that the maximal value of von Neumann is given by
\begin{equation}
    S(\rho_A) = m \ \text{log}_2 d.
\end{equation}
Pure state $|\Phi\rangle$ achieving this limit is called maximally entangled for partition $(A/B)$.
States that are maximally entangled for any possible partition are called AME.

Let us concentrate from now on on the qubits ($d = 2$). It has been shown \cite{HIGUCHI2000213,PhysRevA.77.060304,PhysRevLett.118.200502} that for qubits number of nonequivalent AME states is very limited. They exist only for combined states of two, three, five, and six qubits. For two qubits they are equivalent to the Bell states e.g.
\begin{equation}
    |\Phi_2\rangle = \frac{1}{\sqrt{2}}(|0\rangle|0\rangle + |1\rangle|1\rangle),
\end{equation}
whereas for three qubits they are GHZ-like states e.g.
\begin{equation}
    |\Phi_3\rangle = \frac{1}{\sqrt{2}}(|0\rangle|0\rangle|0\rangle + |1\rangle|1\rangle|1\rangle).
\end{equation}
\begin{figure}[b]
\includegraphics[width=8.4cm]{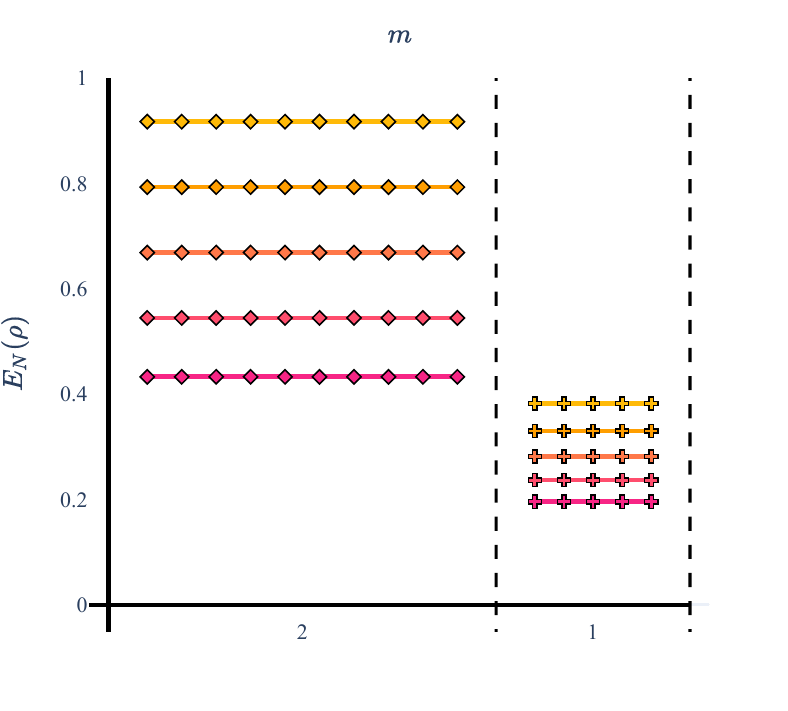}
\caption{Logarithmic negativity of AME state $\ket{\Phi_5}$ defined in Eq. \ref{eq:4} after the action of depolarizing channel. We show results for partitions with $m=1$ and $m=2$, for all possible permutations. The logarithmic negativity decreases monotonically with $p$ for $p = [0.225,0.25,0.275,0.3,0.325]$. Results for given $p$ and $m$ are connected with lines.}
\label{fig1}
\end{figure}
\begin{figure}[t]
\includegraphics[width=8.4cm]{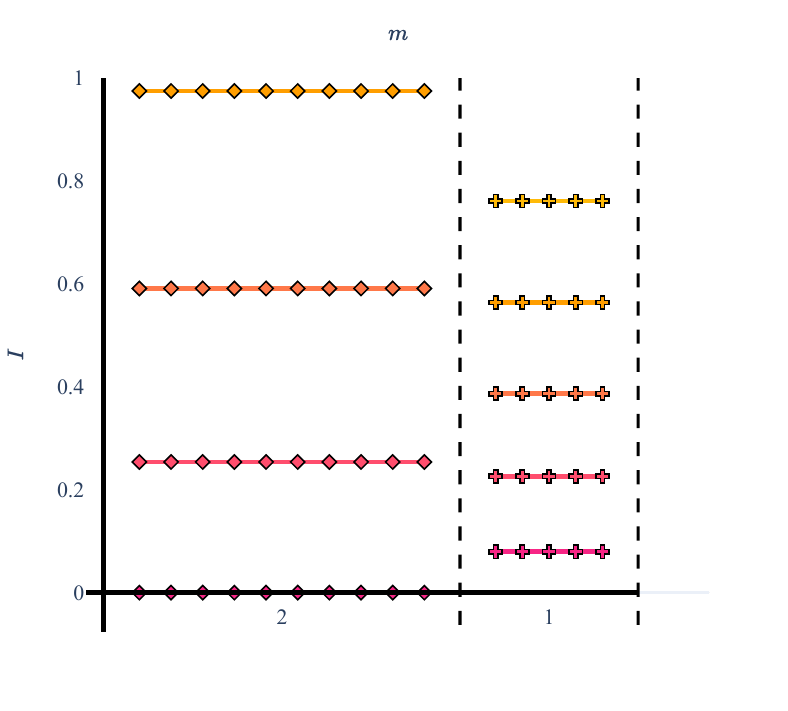}
\caption{Coherent information of AME state $\ket{\Phi_5}$ defined in Eq. \ref{eq:4} after the action of depolarizing channel. We show results for partitions with $m=1$ and $m=2$, for all possible permutations. The coherent information decreases monotonically with $p$ for $p = [0.025,0.05,0.075,0.1,0.125]$. Results for given $p$ and $m$ are connected with lines.}
\label{fig2}
\end{figure}
Exemplary five-qubit AME state are given by
\begin{eqnarray} \label{eq:4}
|0_L\rangle=\frac{1}{4}(|0\rangle|0\rangle|0\rangle|0\rangle |0\rangle
+|1\rangle|0\rangle|0\rangle|1\rangle|0\rangle 
+|0\rangle|1\rangle|0\rangle|0\rangle|1\rangle\nonumber\\
+|1\rangle|0\rangle|1\rangle|0\rangle|0\rangle
+|0\rangle|1\rangle|0\rangle|1\rangle|0\rangle
-|1\rangle|1\rangle|0\rangle|1\rangle|1\rangle\nonumber\\
-|0\rangle|0\rangle|1\rangle|1\rangle|0\rangle
-|1\rangle|1\rangle|0\rangle|0\rangle|0\rangle
-|1\rangle|1\rangle|1\rangle|0\rangle|1\rangle\nonumber\\
-|0\rangle|0\rangle|0\rangle|1\rangle|1\rangle
-|1\rangle|1\rangle|1\rangle|1\rangle|0\rangle
-|0\rangle|1\rangle|1\rangle|1\rangle|1\rangle\nonumber\\
-|1\rangle|0\rangle|0\rangle|0\rangle|1\rangle
-|0\rangle|1\rangle|1\rangle|0\rangle|0\rangle
-|1\rangle|0\rangle|1\rangle|1\rangle|1\rangle\nonumber\\
+|0\rangle|0\rangle|1\rangle|0\rangle|1\rangle)\nonumber\\
\end{eqnarray}
and
\begin{eqnarray}
|1_L\rangle=\frac{1}{4}(|1\rangle|1\rangle|1\rangle|1\rangle |1\rangle
+|0\rangle|1\rangle|1\rangle|0\rangle|1\rangle 
+|1\rangle|0\rangle|1\rangle|1\rangle|0\rangle\nonumber\\
+|0\rangle|1\rangle|0\rangle|1\rangle|1\rangle
+|1\rangle|0\rangle|1\rangle|0\rangle|1\rangle
-|0\rangle|0\rangle|1\rangle|0\rangle|0\rangle\nonumber\\
-|1\rangle|1\rangle|0\rangle|0\rangle|1\rangle
-|0\rangle|0\rangle|1\rangle|1\rangle|1\rangle
-|0\rangle|0\rangle|0\rangle|1\rangle|0\rangle\nonumber\\
-|1\rangle|1\rangle|1\rangle|0\rangle|0\rangle
-|0\rangle|0\rangle|0\rangle|0\rangle|1\rangle
-|1\rangle|0\rangle|0\rangle|0\rangle|0\rangle\nonumber\\
-|0\rangle|1\rangle|1\rangle|1\rangle|0\rangle
-|1\rangle|0\rangle|0\rangle|1\rangle|1\rangle
-|0\rangle|1\rangle|0\rangle|0\rangle|0\rangle\nonumber\\
+|1\rangle|1\rangle|0\rangle|1\rangle|0\rangle)\nonumber\\
\end{eqnarray}
and the six-qubit AME state is given by
\begin{eqnarray}\label{phi6}
|\Phi_6\rangle=\frac{1}{\sqrt{2}}(|0\rangle|0_L\rangle+|1\rangle|1_L\rangle)   
\end{eqnarray}

As stated before there are no AME states of four, seven, or more qubits. Note that presented AME states are only exemplary. For a given number of qubits, other AME state can be generated by  local unitary operations (LUO) of the exemplary ones.

We are interested in how entanglement changes in AME states when each qubit is sent through an identical quantum channel. A quantum channel is a completely positive trace-preserving map given by
\begin{equation}
    \lambda_p(\rho) = \sum_i K_{p,i} \rho K^\dagger_{p,i},
\end{equation}
with Kraus operators $K_{p,i}$ fulfilling
\begin{equation}
    \sum_i K^\dagger_{p,i} K_{p,i} = I
\end{equation}
$\rho$ is single qubit density matrix.
We will use the most popular channels i.e. depolarizing channel
\begin{eqnarray}
\lambda_p (\rho)=(1-\frac{3p}{4})I  \rho I^{\dagger} + \frac{p}{4} X \rho X{^\dagger}+\nonumber\\
+ \frac{p}{4} Y \rho Y{^\dagger} + \frac{p}{4} Z \rho Z{^\dagger}  
\end{eqnarray}
dephasing channel
\begin{eqnarray}
\lambda_p (\rho)=(1-p)I  \rho I^{\dagger} + p |0 \rangle \langle 0| \rho |0\rangle \langle 0|+\nonumber\\
+p |1\rangle \langle 1| \rho |1\rangle \langle 1|  
\end{eqnarray}
where $X$, $Y$ and $Z$ are Pauli matrices.
Equivalently the action of dephasing channel can be written as
\begin{eqnarray} \label{eq13}
\lambda_p (\rho)=(1-\frac{p}{2})I  \rho I^{\dagger} + \frac{p}{2} Z \rho Z^{\dagger}. 
\end{eqnarray}
The final state of the whole set of qubits is given by
\begin{equation}
    \rho' = \Lambda_p(\rho),
\end{equation}
where local, symmetric channel $\Lambda_p$ is
\begin{equation}
    \Lambda_p(\rho) = \left(\bigotimes^n_{j=1} \lambda_p \right
    )\rho.
\end{equation}
Here $\rho$ and $\rho'$ stand for multiqubit density matrices.
We will show that although some AME states (for a given number of qubits $n$) are equivalent with respect to LUO their behavior under identical quantum channels may break this equivalence.
\begin{figure}[b]
\includegraphics[width=8.4cm]{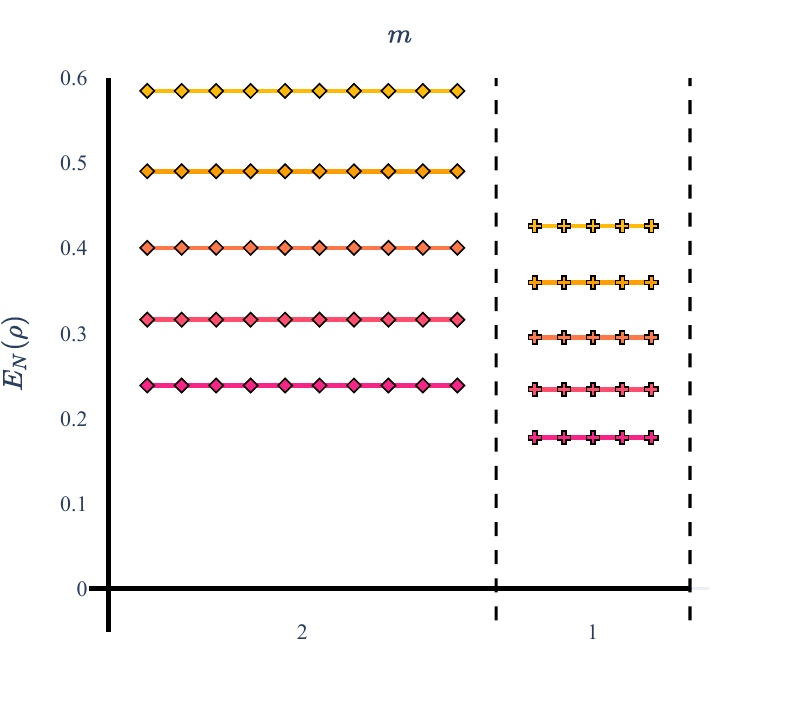}
\caption{Logarithmic negativity of AME state $\ket{\Phi_5}$ defined in Eq. \ref{eq:4} after the action of dephasing channel. We show results for partitions with $m=1$ and $m=2$, for all possible permutations. The logarithmic negativity decreases monotonically with $p$ for $p = [0.5,0.55,0.6,0.65,0.7]$. Results for given $p$ and $m$ are connected with lines.}
\label{fig3}
\end{figure}

\begin{figure}[t]
\includegraphics[width=8.4cm]{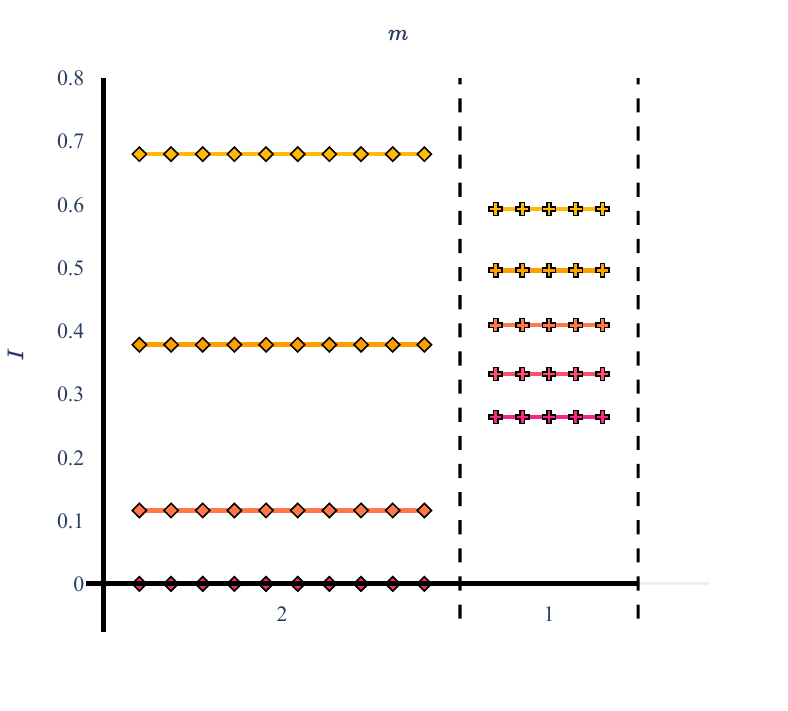}
\caption{Coherent information of AME state $\ket{\Phi_5}$ defined in Eq. \ref{eq:4} after the action of dephasing channel. We show results for partitions with $m=1$ and $m=2$, for all possible permutations. The coherent information decreases monotonically with $p$ for $p = [0.15,0.2,0.25,0.3,0.35]$. Results for given $p$ and $m$ are connected with lines.}
\label{fig4}
\end{figure}
For pedagogical reasons let us first consider the two simple examples of two-qubit AME states
\begin{equation}
    |\Phi_2\rangle = \frac{1}{\sqrt{2}}(|0\rangle|0\rangle+|1\rangle|1\rangle)
\end{equation}
and
\begin{equation}
    |\Phi'_2\rangle = (I\otimes H)\ket{\Phi_2} = \frac{1}{2}(|0\rangle|0\rangle +|0\rangle|1\rangle+|1\rangle|0\rangle-|1\rangle|1\rangle), 
\end{equation}
where we used the Hadamard operation
\begin{equation}
    H=\frac{1}{\sqrt{2}}(X +Z). 
\end{equation}

First, we consider these states undergoing a depolarizing channel. The entanglement change will be identical in both cases because this channel commutes with $(I\otimes H)$ and the entanglement measures are invariant with respect to LUO. 

In the second scenario, the states pass through a dephasing channel. Unlike the depolarizing channel, this channel does not commute with $(I\otimes H)$, leading to different entanglement measures of the final states.

To generalize this argument to any number $n$ of qubits, we first observe that the depolarizing channel commutes with local unitary operations. Consequently, AME states which are equivalent under local unitary operations will experience identical entanglement measures changes, since entanglement is LUO-invariant. However, channels that fail to commute with LUO may exhibit more complex behavior.

For $n>2$ qubits, we fix the number $m$ of qubits in subset $A$ and investigate whether the final entanglement depends on the specific choice of qubits forming $A$. While permutationally invariant states would yield identical entanglement for all $m$-qubit subsets, AME states need not be permutationally invariant. We demonstrate that for certain channels, we can select initial AME states where the entanglement remains identical across different $m$-qubit subsets. 

For the transmitted states, we compute entanglement measures. Practically computable measures include negativity~\cite{eisert1999comparison,PhysRevA.58.883} and logarithmic negativity~\cite{PhysRevLett.95.090503}. We employ logarithmic negativity due to its property as an upper bound on distillable entanglement. Logarithmic negativity is defined as
\begin{equation}
E_N(\rho) = \log_2||\rho^{T_B}||,
\end{equation}
where $T_B$ denotes partial transposition and $||\rho^{T_B}|| = \text{Tr}\sqrt{\rho^{T_B \dagger} \rho^{T_B}}$.

The lower bound on distillable entanglement is given by the maximum of two coherent informations, $I(A>B)$ and $I(B>A)$, where:
\begin{equation}
I(A>B) = S(\rho_B) - S(\rho_{AB}),
\end{equation}
with $I(B>A)$ defined analogously by exchanging $S(\rho_B)$ with $S(\rho_A)$~\cite{devetak2005distillation}. The action of our channels on an $n$-qubit state is of the form
\begin{eqnarray}
    \Lambda^{\otimes n}(\rho_{AB})=\sum_{i} p_{i} U_{Ai} \otimes U_{Bi} \ \rho_{AB} \ U_{Ai}^{\dagger} \otimes U_{Bi}^{\dagger}
\end{eqnarray}
where $U_{Ai}$ and $U_{Ai}$ are tensor products of single-qubit unitary operations. The entropy of the reduced density matrix of subsystem $B$ after the action of the channels satisfies the following chain of relations
\begin{eqnarray}
    S(\Lambda^{\otimes m}(\rho_{B}))=S(\sum_{i} p_{i} U_{Bi} \rho_{B} U_{Bi}^{\dagger}) \nonumber\\
    \geq \sum_{i} p_{i} S(U_{Bi} \rho_{B} U_{Bi}^{\dagger})=S(\rho_{B}) 
\end{eqnarray}
Similar relations hold for the entropy of the reduced density matrix of subsystem $A$ after the action of the channels. At the beginning $S(\rho_B)=S(\rho_A)$ ($|A|\leq |B|$). Entropy of subsystem $A$ is maximal and cannot increase whereas the entropy of subsystem $B$ can increase. Hence we obtain that $I(A>B)$ is greater than $I(B>A)$.

\begin{figure}[b]
\includegraphics[width=8.4cm]{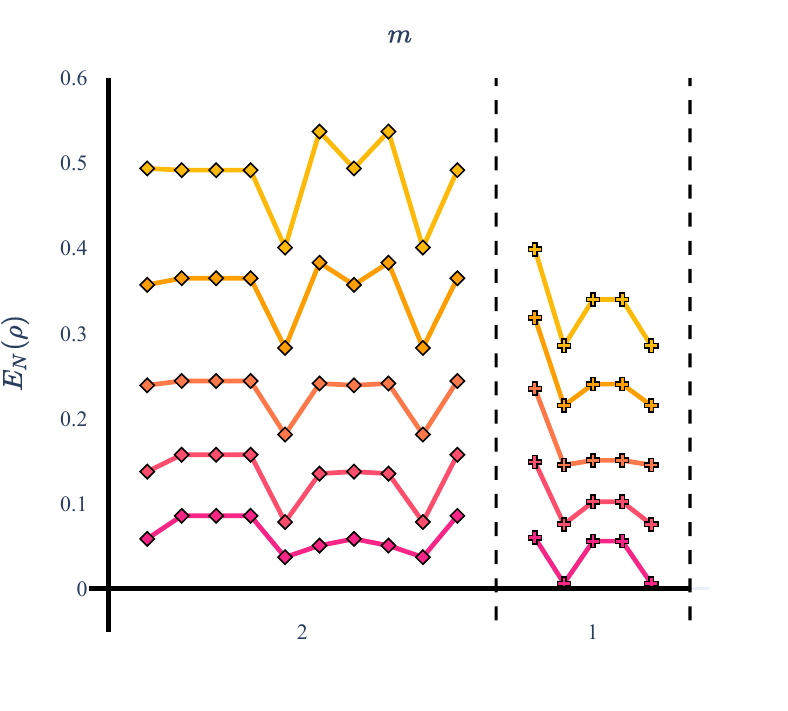}
\caption{Logarithmic negativity of AME state $\ket{\Phi_5'}$ defined in Eq. \ref{prim} after the action of dephasing channel. We show results for partitions with $m=1$ and $m=2$, for all possible permutations. The logarithmic negativity decreases monotonically with $p$ for $p = [0.5,0.55,0.6,0.65,0.7]$. Results for given $p$ and $m$ are connected with lines.}
\label{fig5}
\end{figure}

\begin{figure}[t]
\includegraphics[width=8.4cm]{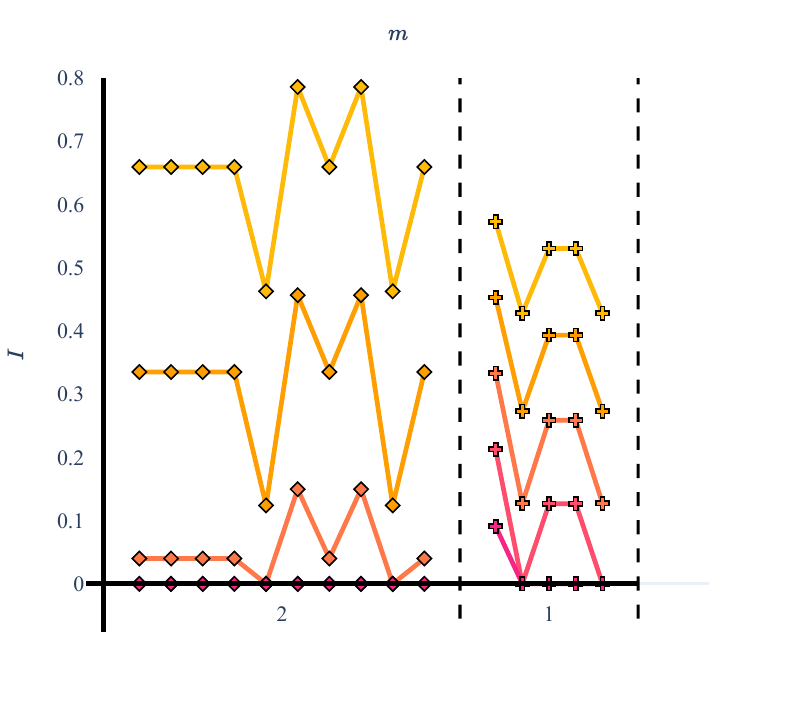}
\caption{Coherent information of AME state $\ket{\Phi_5'}$ defined in Eq. \ref{prim} after the action of dephasing channel. We show results for partitions with $m=1$ and $m=2$, for all possible permutations. The coherent information decreases monotonically with $p$ for $p = [0.15,0.2,0.25,0.3,0.35]$. Results for given $p$ and $m$ are connected with lines.}
\label{fig6}
\end{figure}
\section{Results} Let us first choose a five-qubit state $|\Phi_5 \rangle = \ket{0_L}$ and depolarizing channel. In Fig. 1 we present how logarithmic negativity and in Fig. 2 coherent information depend both on $p$ and the choice of qubits in set $A$. In presented figures we show results for $m=1$ and $m=2$ (which are the only possible ones for $n=5$). We see that in both cases entanglement does not depend on which qubits form the set $A$. Moreover, as mentioned previously it does not depend on a choice of AME state from a class of states that are related by local unitary operations. Not only LUO but also local in the sense of partition $U_A \otimes U_B$.

Next let us shift our attention to the dephasing channel. 
In Fig. 3 we present how logarithmic negativity and in Fig. 4 coherent information depend both on $p$ and the choice of a qubits in set $A$. In presented figures we show results for $m=1$ and $m=2$. We observe the same behavior as before. Entanglement does not depend on which qubits are in the set $A$. However, now we cannot argue as before that entanglement does not depend on a choice of AME state. So let us check what happens when we choose other five qubit AME state related by local unitary operations to the state defined in Eq. \ref{eq:4}. Namely

\begin{equation}\label{prim}
    \ket{\Phi_5'} = (I\otimes I\otimes I\otimes I\otimes H)\ket{\Phi_5}.
\end{equation}
In Fig. 5 we present how logarithmic negativity and in Fig. 6 coherent information depend both on $p$ and the choice of a qubits in set $A$. In presented figures we show results for $m=1$ and $m=2$. Now we see that results depend on particular choice of qubits. To emphasize our observation that some AME states maintain the symmetry present in the entanglement measures under noise while others do not, we present in Fig. 7 the comparison of logarithmic negativity for states $\ket{\Phi_5}$ and $\ket{\Phi_5'}$ after the action of the dephasing channels with $p=0.36$.

We can generalize these results to other channels. For this let us consider the following channel (so-called Pauli channel \cite{Bengtsson_Zyczkowski_2006})
\begin{eqnarray}
    \Lambda (\rho) = (1-p-q-r)I \rho I+pX \rho X+\nonumber\\
    +qY \rho Y +rZ \rho Z
\end{eqnarray}
where, $p,q,r \geq 0$ and $p+q+r \leq 1$. If we choose $p=q\neq r$ we obtain that for state $\ket{\Phi_5}$ entanglement does not depend on particular choice of qubits. However, for the state $\ket{\Phi_5'}$ the entanglement does depend on the choice of qubits. Remember that for depolarizing channel ($p=q=r$) entanglement still does no depend on the choice of qubits. For six-qubit state of  Eq. \ref{phi6} for depolarizing and dephasing channels we observe a similar behavior as before, i.e. entanglement does not depend on the choice of qubits.
\begin{figure}[t]
\includegraphics[width=8.4cm]{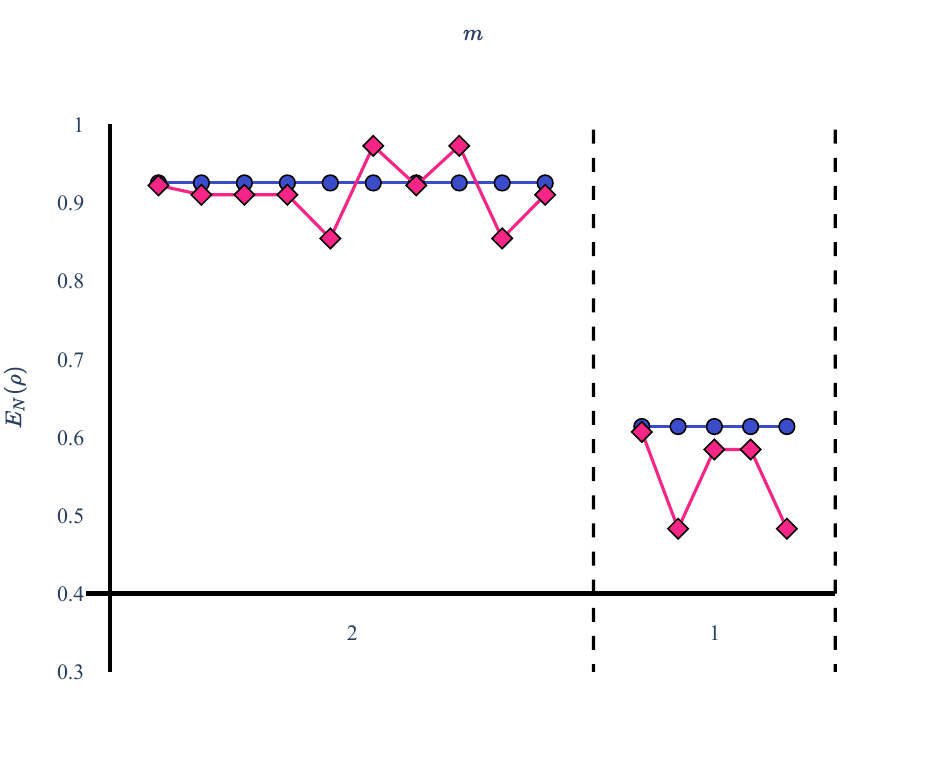}
\caption{Logarithmic negativity of AME state $\ket{\Phi_5}$ (blue) together with $\ket{\Phi_5'}$ (pink) after the action of dephasing channel with $p=0.36$. We show results for partitions with $m=1$ and $m=2$, for all possible permutations. Results for given $m$ are connected with lines.}
\label{fig1}
\end{figure}
\section{Conclusions} In this work, we explored the impact of noisy quantum channels on the entanglement properties of absolutely maximally entangled states. By analyzing logarithmic negativity and coherent information, we observed that the depolarizing channel preserves the symmetry of entanglement across all local unitary-equivalent AME states, while the dephasing channel and certain Pauli channels can break this symmetry, leading to state-dependent entanglement behavior. Our results underscore the importance of selecting appropriate AME states and channels for quantum communication protocols, as not all states exhibit the same resilience to noise. These findings contribute to a deeper understanding of entanglement dynamics in noisy environments and offer practical guidance for optimizing quantum information processing tasks involving AME states. Future research could extend this analysis to higher-dimensional systems or explore the implications for entanglement distillation protocols.

{\it Acknowledgements.} AG, and AW were supported by the National Science Centre (NCN, Poland) under the Maestro Grant no. DEC-2019/34/A/ST2/00081. JW was supported by the National Science Centre (NCN, Poland) within the Chist-ERA project (Grant No. 2023/05/Y/ST2/00005).

\bibliographystyle{apsrev4-2}

\end{document}